\documentclass{aa}
\usepackage{graphicx}
\usepackage{txfonts}

\newcommand{\ls}    {\object{LS~5039}}
\newcommand{\rx}    {\object{RX~J1826.2$-$1450}}
\newcommand{\lsrx}  {\object{LS~5039}/\object{RX~J1826.2$-$1450}}

\def\simless{\mathbin{\lower 3pt\hbox
     {$\rlap{\raise 5pt\hbox{$\char'074$}}\mathchar"7218$}}}   %< or of order
\def\simmore{\mathbin{\lower 3pt\hbox
     {$\rlap{\raise 5pt\hbox{$\char'076$}}\mathchar"7218$}}}   %> or of order

\def\msun{~{\rm M}_\odot}
\def\rsun{~{\rm R}_\odot}

\begin{document}

%\thesaurus{08(08.09.2 \object{RX~J1826.2$-$1450}; 08.09.2
%\object{LS~5039}; 08.09.2 \object{NVSS J182614$-$145054}; 13.25.5;
%08.22.3; 13.18.5)}

\title{Long-term X-ray variability of the
microquasar system LS~5039/RX~J1826.2$-$1450}

\subtitle{}

\author{P. Reig\inst{1}
\and M. Rib\'o\inst{2}
\and J.~M. Paredes\inst{2,}\thanks{CER on Astrophysics, Particle Physics
and Cosmology. Universitat de Barcelona}
\and J. Mart\'{\i}\inst{4}
}

\institute{
G.A.C.E, Institut de Ci\`encies dels Materials, Universitat de Valencia, 46071 Paterna-Valencia, Spain\\
\email{pablo.reig@uv.es}
\and Departament d'Astronomia i Meteorologia, Universitat de Barcelona, Av. Diagonal 647, 08028 Barcelona, Spain\\
\email{mribo@am.ub.es; josep@am.ub.es}
\and Departamento de F\'{\i}sica, Escuela Polit\'ecnica Superior, Universidad de Ja\'en, Virgen de la Cabeza 2, 23071 Ja\'en, Spain\\
\email{jmarti@ujaen.es}
}

\authorrunning{Reig et~al.}
\titlerunning{X-ray variability of LS~5039/RX~J1826.2$-$1450}
%\titlerunning{X-ray variability of LS~5039/RX~J1826.2$-$1450}

\offprints{P. Reig, \\ \email{pablo.reig@uv.es}}

\date{Received / Accepted}

\abstract{
We report on the results of the spectral and timing analysis of a BeppoSAX
observation of the microquasar system \lsrx. The source was found in a
low-flux state with $F_{\rm X}$(1--10~keV)= $4.7 \times 10^{-12}$
erg~cm$^{-2}$~s$^{-1}$, which represents almost one order of magnitude lower
than a previous RXTE observation 2.5 years before. The 0.1--10~keV spectrum is
described by an absorbed power-law continuum with photon-number spectral index
$\Gamma=1.8\pm0.2$ and hydrogen column density of $N_{\rm H}=1.0^{+0.4}_{-0.3}
\times 10^{22}$~cm$^{-2}$. According to the orbital parameters of the system
the BeppoSAX observation covers the time of an X-ray eclipse should one occur.
However, the 1.6--10~keV light curve does not show evidence for such an event,
which allows us to give an upper limit to the inclination of the system. The
low X-ray flux detected during this observation is interpreted as a decrease
in the mass accretion rate onto the compact object due to a decrease in the
mass-loss rate from the primary.
}

\maketitle

\section{Introduction} \label{introduction}

\lsrx\ was first identified as a new massive X-ray binary by Motch et~al.
(\cite{motch97}). Interest in this source has grown significantly because it
has turned out to be a source of persistent radio-emitting relativistic jets,
and because it is believed to be physically associated with the $\gamma$-ray
source \object{3EG~J1824$-$1514} (Paredes et~al. \cite{paredes00},
\cite{paredes02}).

The first radio detection was reported by Mart\'{\i} et~al. (\cite{marti98})
using the Very Large Array, but the discovery of radio jets was only possible
when the source was observed at milliarcsecond scales with the Very Long
Baseline Array (Paredes et~al. \cite{paredes00}). The radio emission is
persistent, non-thermal and variable, but no strong radio outbursts or
periodic variability have been detected (Rib\'o et~al. \cite{ribo99},
hereafter R99; Rib\'o \cite{ribo02t}).

In the optical band \lsrx\ appears as a bright $V$=11.2, O6.5V((f)) star
showing little variability on timescales of months to years (Clark et~al.
\cite{clark01}). Variations of $\sim$0.4~mag have been reported in the
infrared ($H$ and $K$ bands) but no obvious mechanisms for such variability
have been proposed (Clark et~al. \cite{clark01}). Recent studies show that
\ls\ is a runaway system moving away from the Galactic plane with a total
systemic velocity of $\sim$150~km~s$^{-1}$ and a component perpendicular to
the Galactic plane larger than 100~km~s$^{-1}$ (Rib\'o et~al. \cite{ribo02};
McSwain \& Gies \cite{mcswain02}). The orbit of \lsrx\ was first studied by
McSwain et~al. (\cite{mcswain01}), who found that \lsrx\ is a short-period,
$P_{\rm orb}=4.117\pm0.011$ day and highly eccentric, $e=0.41\pm0.05$, system.
Spectroscopic observations carried out in July 2002 by Casares et~al.
(\cite{casares03}) seem to confirm the orbital period of the system, but
neither confirm nor reject the proposed eccentricity.

%------------------------------------------------------------------------------
\begin{figure}[htpb]
\resizebox{\hsize}{!}{\includegraphics{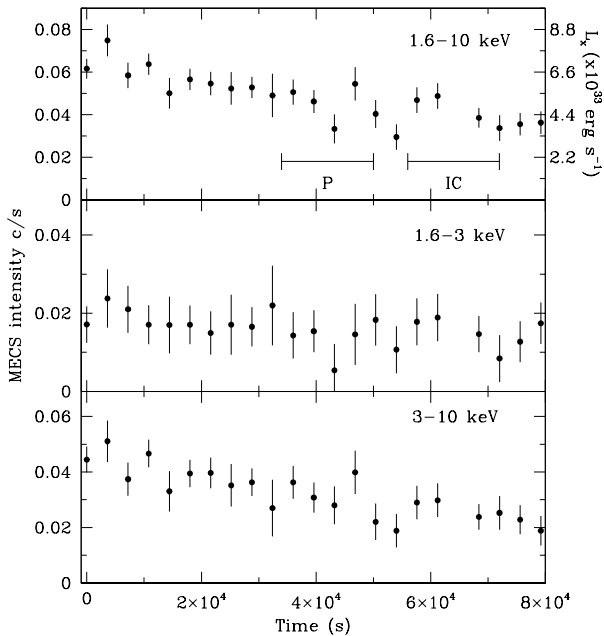}}
\caption[]{MECS background subtracted light curves of \object{RX~J1826.2$-$1450}
 in different energy ranges. Each point represents 60 minutes. Time zero 
 corresponds to JD\,2,451,825.75. The expected time of periastron (P) and 
 inferior conjunction (IC) of the primary are indicated in the top panel,
 where the error bars represent the uncertainty in the zero point of the 
 ephemeris. The time difference between these two points is 
 2.2 $\times$ 10$^4$ seconds. An estimate of the unabsorbed X-ray luminosity 
 according to the model of Table~\ref{specfit} is also given in the right
 axis of the top panel.}
\label{lc}
\end{figure}
%------------------------------------------------------------------------------

In the X-rays \lsrx\ has been studied by R99 using RXTE data. The X-ray timing
analysis indicates the absence of pulsed or periodic emission on time scales
of 0.02--2000 seconds. The source spectrum is well represented by a power-law
model, plus a Gaussian component describing a strong iron line at 6.6~keV.
Significant emission is seen up to 30~keV, and no exponential cut-off at high
energy is required. The X-ray luminosity has been shown to be compatible with
accretion from the stellar wind of the optical companion (McSwain \& Gies
\cite{mcswain02}).

One of the open questions that remains to be solved is the nature of the
compact object of the system as no conclusive data exists. The absence of
X-ray pulsations and exponential cut-off at $\sim$30~keV and the persistent
non-thermal emission favours the black-hole classification (R99). On the other
hand, the value of the mass function of the system $f(m)=0.00103\msun$
(McSwain et~al. \cite{mcswain01}) and the characteristics of its accretion
mechanism (McSwain \& Gies \cite{mcswain02}) appear to be more consistent with
a neutron star companion. 

In this work we present an analysis of the X-ray timing and spectroscopic
properties of \rx\ using the narrow field instruments on board the BeppoSAX
satellite. This observation permits us to derive a more accurate value of the
absorption to the source and study its long-term X-ray variability. We also
present optical spectroscopic observations around the H$\alpha$ line carried
out at the Skinakas Observatory.

%------------------------------------------------------------------------------
\begin{figure}[htpb]
\resizebox{\hsize}{!}{\includegraphics{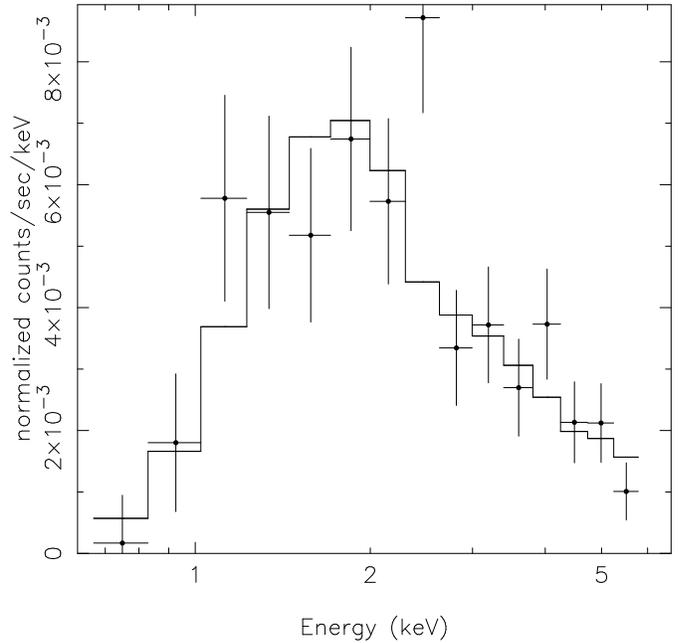}}
\caption[]{LECS energy spectrum of \object{RX~J1826.2$-$1450} and the 
best-fit model displayed in Table~\ref{specfit}.}
\label{lecs}
\end{figure}
%------------------------------------------------------------------------------

\section{Observations and results} \label{observations}

\subsection{X-ray observations}

\lsrx\ was observed by BeppoSAX on October 8, 2000 (JD 2451825.79--2451826.71)
for about 80~ks. BeppoSAX  (Boella et~al. \cite{boella97a}) carried four
narrow field instruments (LECS, MECS, HPGSPC and PDS) and a wide field camera
(WFC). Here we report on the results from the lower energy detectors: the
Low-Energy (LECS) and Medium-Energy (MECS) Concentrator Spectrometers. They
consisted of a set of four and three X-ray concentrators respectively together
with imaging gas scintillation proportional counter detectors located at the
focal planes. In the LECS (Parmar et~al. \cite{parmar97}), one of these
detectors was sensitive to X-rays in the energy range 0.1--10~keV, while the
other three covered an energy range of 1.3--10~keV. The MECS (Boella et~al.
\cite{boella97b}) operated in the energy range 1.6--10~keV. Each detector unit
provided a total collecting (geometrical) area of $\sim$124~cm$^2$, an energy
resolution of $<$~8.8~\% at 6~keV and a maximum time resolution of 16$\mu$s.
Note that one of the MECS became inoperative on May 9, 1997 due to a fault in
the unit's gas cell high voltage supply.

Light curves and spectra were extracted from circular regions with a radius of
4\arcmin\ for both the LECS and the MECS instruments.  We selected two
source-free regions (of radius 4\arcmin\ each) in the field of view, and
obtained a mean light curve in order to account for the instrumental
background. For the spectral analysis, the background was obtained from a
blank field using a region similar in size and position to the source
extraction region.

%------------------------------------------------------------------------------
\begin{figure}[htpb]
\resizebox{\hsize}{!}{\includegraphics{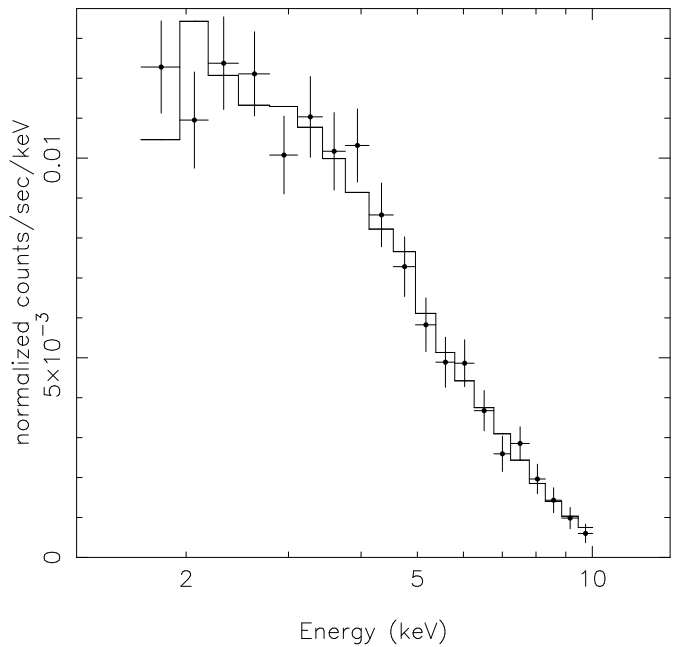}}
\caption[]{MECS energy spectrum of \object{RX~J1826.2$-$1450} and the best-fit
 model displayed in Table~\ref{specfit}.}
\label{mecs}
\end{figure}
%------------------------------------------------------------------------------

The X-ray intensity in the energy range 1.6--10~keV decreased with time from a
value of $\sim$0.06 MECS count~s$^{-1}$ at the beginning of the observation to
$\sim$0.03 MECS count~s$^{-1}$ at the end (Fig.~\ref{lc}). A fit to a constant
gave an unacceptable reduced $\chi^2$ ($>$2). However, adding a linear term
with negative slope we obtained a reduced $\chi^2 \sim$1. An F-test gave a
probability of less than 0.7\% that this improvement in the quality of the fit
could occur by chance. The flux variability occurs at higher energies. Below
$\sim$3~keV the flux remains fairly constant (Fig.~\ref{lc}).

Acceptable fits of the X-ray energy spectra were obtained with an absorbed
power-law model for both LECS (Fig.~\ref{lecs}) and MECS
(Fig.~\ref{mecs}).  Table~\ref{specfit} shows the best-fit spectral
parameters. The power law was modified by a low-energy absorption
component of the form $e^{-N_{\rm H} \sigma(E)}$, where $N_{\rm H}$ is the
equivalent hydrogen column density and $\sigma(E)$ the photoelectric
cross-section of Morrison \& McCammon (\cite{mm83}). The power-law index
and the hydrogen column density have values of $\sim$1.8 and $\sim$ 1.0
$\times$ 10$^{22}$~cm$^{-2}$, respectively  (see discussion in
Sect.~\ref{hydrogen}). The LECS (0.1--5~keV) and MECS (1.6--10~keV) 
absorbed X-ray luminosity was $1.9 \times 10^{33}$ erg~s$^{-1}$ and $4.6
\times 10^{33}$ erg~s$^{-1}$, respectively, assuming a distance of 2.9~kpc
(Rib\'o et~al. \cite{ribo02}).

\subsection{Optical observations}

Optical spectroscopic observations were made with the 1.3\,m telescope of the
Skinakas Observatory in Crete (Greece) on the night of July 26, 1999 and
September 10, 2002. The instrumental set-up consisted of a 2000$\times$800 ISA
SITe CCD and a 1302 l~mm$^{-1}$ grating, giving a dispersion of $\sim$1
\AA/pixel. The reduction of the spectra was made using the STARLINK {\em
Figaro} package (Shortridge et~al. \cite{shortridge01}), while their analysis
was performed using the STARLINK {\em Dipso} package (Howarth et~al.
\cite{howarth98}).

The wavelength coverage of the observations included the H$\alpha$ line
($\lambda6563$ \AA). The strength of the stellar wind emission in H$\alpha$
provides information about the mass-loss rate in O-type stars (Puls et~al.
\cite{puls96} and references therein). The measurement of the equivalent width
of the H$\alpha$ line (EW(H$\alpha$)) in \ls\ is hampered by the presence of
other stellar lines, especially \ion{He}{ii} $\lambda$6527. Nevertheless, when
the emission is low the main factor contributing to the uncertainty in the
equivalent width is the difficulty in obtaining the intensity of the
continuum. In Table~\ref{ew} we give the values of EW(H$\alpha$) as
contemporaneous as possible with the X-ray measurements. 

%The errors are the standard deviation of ten measurements corresponding to
%ten choices of the spectral continuum. 

\section{Discussion} \label{discussion}

\subsection{Hydrogen column density} \label{hydrogen}

The RXTE observation reported by R99 was not able to provide a reliable value
of the absorption to the source due to the insensitivity of the detectors to
X-rays below 2.5~keV, where most of the interstellar absorption of X-ray
photons takes place. The sensitivity of the BeppoSAX narrow field instruments
to X-rays below 2~keV, especially LECS, allows us to derive a more accurate
value of the hydrogen column density, $N_{\rm H}=1.0^{+0.4}_{-0.3} \times
10^{22}$~cm$^{-2}$.  Assuming $E(B-V)=A_V/3.1$ (Rieke \& Lebofsky
\cite{rieke85}) the X-ray absorption can be converted into optical extinction
by means of the empirical relations of Predehl \& Schmitt (\cite{predehl95}),
to obtain $E(B-V)=1.8^{+0.7}_{-0.5}$, or those of Ryter et~al.
(\cite{ryter75}) and Gorenstein (\cite{gorenstein75}), to obtain
$E(B-V)=1.5^{+0.6}_{-0.4}$.  These values of $E(B-V)$ are considerably higher
than the unreliable value of $E(B-V)\approx 0.4$ derived from the PCA/RXTE
observation (R99). 

On the other hand, optical photometry of \ls\ can be found in Drilling
(\cite{drilling91}) and Lahulla \& Hilton (\cite{lahulla92}) who gave a value
for the colour $(B-V)=0.95\pm0.01$ and Clark et~al. (\cite{clark01}), who
reported $(B-V)=0.85\pm0.02$. Using the average of these values and assuming
an intrinsic colour index of $(B-V)_0=-0.30\pm0.02$ for an O6.5V star (Lejeune
\& Scharer \cite{lejeune01}) we obtain a colour excess of $E(B-V)=1.2\pm0.1$.
The extinction obtained from the  BeppoSAX observation is somehow higher
than (although statistically consistent with) that derived from the optical
observations. One explanation for this difference might be the fact that
the hydrogen column density is more sensitive to absorption by both
interstellar and intrinsic (i.e. in the vicinity of the X-ray source) matter,
while the optical indicators would mainly account for the interstellar
absorption. The presence of significant amounts of cold matter in the
surroundings of the system is supported by the detection of the relatively
strong iron (equivalent width of 0.75$\pm$0.06~keV) emission line at 6.6~keV
in the 3--30~keV X-ray spectrum of \lsrx\ (R99). Unfortunately, due to limited
signal-to-noise ratio no line is detected in the BeppoSAX data.

%------------------------------------------------------------------------------
\begin{table}
\begin{center}
\caption{\label{fit} Spectral fit results for the absorbed power-law model. 
Uncertainties are given 68\% confidence for one parameter of interest.}
\label{specfit}
\begin{tabular}{ll}
\hline \hline \noalign{\smallskip}
Parameters                         & Best-fit values \\
\noalign{\smallskip} \hline \noalign{\smallskip}
\multicolumn{2}{c}{LECS} \\
\noalign{\smallskip} \hline \noalign{\smallskip}
%Energy range (keV)              & 0.1--5.0 \\
$N_{\rm H}$ (cm$^{-2}$)         & $1.0^{+0.4}_{-0.3}$ $\times 10^{22}$ \\
$\Gamma$                        & 1.8$\pm$0.3 \\
$F_{\rm X}$ (0.1--5~keV) (erg cm$^{-2}$ s$^{-1}$) & $1.9 \times 10^{-12}$ \\
$\chi^2_{\rm r}$ (dof)          & 1.25 (12)     \\
\noalign{\smallskip} \hline \noalign{\smallskip}
\multicolumn{2}{c}{MECS} \\
\noalign{\smallskip} \hline \noalign{\smallskip}
%Energy range (keV)              & 1.6--10 \\
$N_{\rm H}$ (cm$^{-2}$)         & $1.2\pm0.4$ $\times 10^{22}$ \\
$\Gamma$                        & 1.8$\pm$0.1 \\
$F_{\rm X}$ (1.6--10~keV) (erg cm$^{-2}$ s$^{-1}$) & $4.6 \times 10^{-12}$ \\
$\chi^2_{\rm r}$ (dof)          & 0.87 (17) \\
\noalign{\smallskip} \hline
\end{tabular}
\end{center}
\end{table}
%------------------------------------------------------------------------------

\subsection{Absence of X-ray eclipse} \label{eclipse}

Using the ephemeris of McSwain et~al. (\cite{mcswain01}), the BeppoSAX
observation covers the orbital phase from 0.89 to 0.11. Note that this
observation took place just four days after the periastron date used by
these authors in their ephemeris, i.e., JD\,2,451,822.12$\pm$0.09.
Consequently, our determination of the orbital phase is not strongly
affected by the lack of accuracy in the value of the orbital period
($P_{\rm orb}=4.117\pm0.011$ days). Hence, periastron took place on
JD\,2,451,826.24$\pm$0.09, or at $4.2\pm0.8\times10^4$~s in Fig.~\ref{lc}.
The epoch of inferior conjunction of the optical star, which would
correspond to the time of an X-ray eclipse, is 0.25 days after periastron
or at orbital phase 0.06 (McSwain et~al. \cite{mcswain01}), which
corresponds to JD\,2,451,826.49$\pm$0.09, or to $6.4\pm0.8\times10^4$~s in
Fig.~\ref{lc}. The duration of a possible eclipse would last less than 0.2
days ($1.7\times10^4$~s) for $M_{\rm opt}=40\msun$, $R_{\rm opt}=10\rsun$
and $i=90\degr$ (shorter for a lower inclination and an eventual lower
mass for the primary) and hence it should have been detected by BeppoSAX. 

Although the light curve does not show statistical significant evidence for an
eclipse, the variability increases between periastron and inferior
conjunction, possibly indicating some kind of distortion of the stellar wind
when the compact star finds itself at minimum distance from the primary.
There is also a trend for the count rate to decrease as the source approaches
inferior conjunction. This trend is detected at energies above 3~keV, but not
at lower energies, contrary to what would be expected if this was due to a
change in absorption, as it happens in \object{Cygnus~X-1} (Wen et~al.
\cite{wen99}). Therefore, this decrease in the hard X-ray flux during the
BeppoSAX observation is probably intrinsic, and may reflect changes in the
accretion rate due to orbital motion. 

The lack of eclipses allows us to set an upper limit on the inclination of the
system $i$. This implies, in the case of a spherical companion  (no
photometric ellipsoidal modulation has been found to date), that $\cos i >
R_{\rm opt}/r$, where $R_{\rm opt}$ is the radius of the optical companion an
$r$ is its distance to the compact object at inferior conjunction (phase
0.06). Assuming a mass of 40$\msun$ and a radius of 10$\rsun$ for an O6.5V
star (Howarth \& Prinja \cite{howarth89}), the canonical mass for a neutron
star (1.4$\msun$) and the orbital parameters of McSwain et~al.
(\cite{mcswain01}), $e=0.41\pm0.05$ and $P_{\rm orb}=4.117\pm0.011$ days, we
obtain an upper limit for the inclination of the system of $i<66\pm2\degr$,
where the upper and lower values correspond to the range of eccentricities
given above.  This inclination is consistent with the one, $i_f$, derived
from the mass function, $f(m)=0.00103\pm0.00020\msun$ (McSwain et~al.
\cite{mcswain01}), for the same parameters, namely, $i_f=60\degr$.

Note that OB stars in X-ray binary systems may be less massive than the
corresponding field stars of the same spectral type (Kaper \cite{kaper01}). A
decrease of the mass of the optical companion by about 10\% reduces $i_f$ to
$54\degr$, while the upper limit hardly changes. On the other hand, assuming
$M_{\rm opt}=40\msun$ and the above mentioned orbital parameters Rib\'o et~al.
(\cite{ribo02}) derived an upper limit for the mass of the compact object of 9
$\msun$ (in order to avoid breackup speed of the primary once $V_{\rm rot}\sin
i$ is known). For a $9\msun$ black-hole companion a similar analysis yields
$i<68\pm2\degr$ and $i_f=8.5\degr$. Similar values are obtained if we decrease
the mass of the optical companion by about 10\%. In either case, the lack of
X-ray eclipses constrains the inclination of the system to values smaller than
70\degr, which is clearly compatible with those found using the mass function
of the system.

%------------------------------------------------------------------------------
\begin{table}
\begin{center}
\caption{X-ray flux and H$\alpha$ equivalent width measurements of \ls.
Errors in $EW$(H$\alpha$) are $\simless 10$\%.}
\label{ew}
\begin{tabular}{l@{~~~}l@{~~~}c@{~~}|c@{~~~}l}
\hline \hline \noalign{\smallskip}
Mission  & Date         & $F$$^a_{\rm X}$ & $EW$(H$\alpha$) & Date \\
         & X-ray obs.   &                 & (\AA)           & Opt. obs. \\
\noalign{\smallskip} \hline \noalign{\smallskip}
RXTE     & \,~8/02/1998 & 40$^b$~~\,      & 2.2$^c$         & Aug 1998 \\
ASCA     & \,~4/10/1999 & 10.9$^b$        & 2.8$^b$         & 26/7/1999 \\
BeppoSAX & \,~8/10/2000 & ~\,4.9$^b$      & 3.1$^c$         & Oct 2000 \\
CHANDRA  & 10/09/2002   & \,10.0$^d$      & 2.8$^b$         & 10/9/2002 \\
\noalign{\smallskip} \hline
\end{tabular}
\begin{list}{}{}
\item[$^a$] $\times 10^{-12}$ erg cm$^{-2}$ s$^{-1}$ in the energy range 0.3--10~keV.
\item[$^b$] This work.
\item[$^c$] McSwain \& Gies (\cite{mcswain02}).
\item[$^d$] Miller (private communication).
\end{list}
\end{center}
\end{table}
%------------------------------------------------------------------------------

\subsection{Long-term X-ray variability} \label{long}

In order to study the long-term X-ray variability of \lsrx\ we obtained the
X-ray luminosity in the same energy ranges as given in previous missions. The
value of the ROSAT 1996 October 0.1--2.4~keV (Motch et~al. \cite{motch97}) and
the RXTE 1998 February 3--30~keV (R99) X-ray luminosities are 7.1 $\times$
10$^{33}$ erg~s$^{-1}$ and 5.2 $\times$ 10$^{34}$ erg~s$^{-1}$, respectively.
The 0.1--2.4~keV X-ray luminosity obtained with LECS is 6.6 $\times$ 10$^{32}$
erg~s$^{-1}$ and that obtained with MECS in the energy range 3--30~keV is 8.0
$\times$ 10$^{33}$ erg~s$^{-1}$ (derived assuming the model given in
Table~\ref{specfit}). This observation took place in 2000 October. All values
are for an assumed distance of 2.9~kpc (Rib\'o et~al. \cite{ribo02}). Clearly
the source was in a low-flux state during the BeppoSAX observation (about one
order of magnitude fainter) when compared to the epoch of the ROSAT and RXTE
observations. As explained above this low flux, however, cannot be accounted
for by invoking a partial X-ray eclipse. We have also inspected archival X-ray
data and found an ASCA observation of \lsrx\ made on 1999 October 4. We have
obtained the X-ray flux in the energy range 0.3--10~keV, shown in
Table~\ref{ew}, which is between the RXTE and BeppoSAX measurements. In
addition, a similar X-ray flux has been recently detected with CHANDRA
(Miller, private communication), and is also reported in Table~\ref{ew}.

We also studied the long-term optical variability of the primary by
measuring the strength of the H$\alpha$ line. Table~\ref{ew} shows the
values of the H$\alpha$ equivalent width together with the X-ray flux from
the missions that have observed \lsrx. The H$\alpha$ emission was weakest
during the BeppoSAX observation in October 2000 and strongest a few months
after the RXTE pointing in February 1998, when the X-ray emission was also
strong (since the line is in absorption, the smaller $EW$(H$\alpha$) the
stronger the emission). Moreover, very similar $EW$(H$\alpha$) are found
contemporaneous to the ASCA and CHANDRA observations, as the detected
X-ray fluxes are, and also between the $EW$(H$\alpha$) measurements
contemporaneous to the RXTE and BeppoSAX observations. Hence, there is a
trend in the sense that the lower the emission in H$\alpha$, the lower the
observed X-ray flux. All these results suggest that a decrease in the
mass-loss rate from the primary induces a decrease in the emitted X-ray
flux of the binary system. In this context, we propose that changes in the
mass-loss rate from the primary, reflected in the change of
$EW$(H$\alpha$), can account for a change in the mass accretion rate onto
the compact object, and hence explain the detected changes in X-ray
luminosity. 

On the other hand, since this binary system has a high eccentricity, we cannot
rule out {\em a priori} that the detected variations in the X-ray flux are not
due to orbital variability. If the mass transfer in \lsrx\ occurs via the
strong stellar wind of the optical primary then, given the high eccentricity
of the system, some variations in the X-ray luminosity with the orbital phase
should be expected. The fraction of the stellar wind captured by the compact
object depends on its velocity relative to the wind. Also, variations in the
density of the wind may produce changes in the amount of absorption material
throughout the orbit giving rise to the X-ray flux variations. Orbital flux
variations have been seen in many high-mass X-ray binaries powered by stellar
wind accretion: \object{4U~1700$-$37} (Haberl \& Day \cite{haberl92}),
\object{GX~301$-$2} (Haberl \cite{haberl91}), \object{Vela~X-1} (Haberl \&
White \cite{haberl90}), \object{4U~1907+09} (Marshall \& Ricketts
\cite{marshall80}). With its small orbit and early-type optical primary \lsrx\
is expected to share many similarities with these systems. 

\begin{figure}[t]
\resizebox{\hsize}{!}{\includegraphics[angle=0]{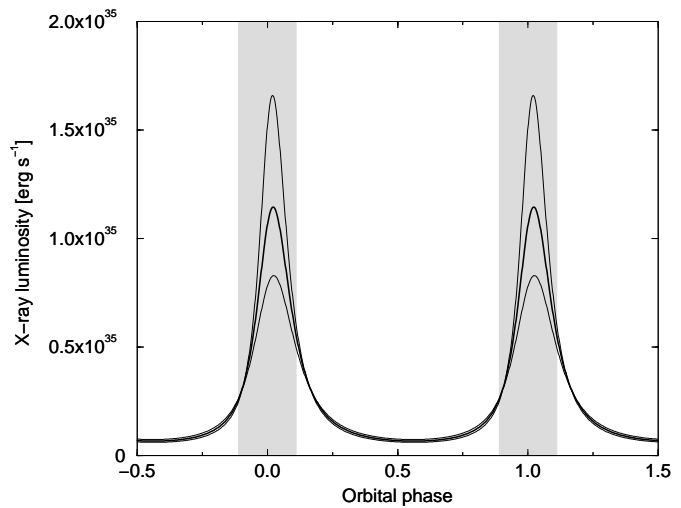}}
\caption[]{Expected variability of the X-ray luminosity in \object{LS~5039}
due to the significantly eccentric orbit of the compact companion around 
the O6.5V((f)) star with a period $P_{\rm orb}=4.117$~d, using a spherically 
symmetrical Bondi-Hoyle accretion model and the parameters quoted in the
text. The thick line corresponds to $e=0.41$, the nominal eccentricity value 
in the McSwain et~al. (\cite{mcswain01}) solution, while the lower and upper 
thin lines are for $e=0.36$ and $e=0.46$, which correspond to the error limits 
quoted by the authors. Phase 0 corresponds to periastron, while the shaded areas correspond to the phase interval covered by our BeppoSAX observation.}
\label{bondi}
\end{figure}

However, a purely Bondi-Hoyle accretion model (Bondi \cite{bondi52}) does
not reproduce the observations. Figure~\ref{bondi} shows the expected
variability of the X-ray luminosity in \lsrx\ using a spherically symmetrical
Bondi-Hoyle accretion model. For each orbital phase and using the formulae
given in the Appendix, we computed the binary separation, the orbital and wind
velocities, the relative velocity, the wind density, the accretion rate and
finally the X-ray luminosity. We note that this is a bolometric  unabsorbed
X-ray luminosity, to be compared to the much lower observed luminosity in the
range 1.6--10~keV (top panel of Fig.~\ref{lc}). Although this discrepancy
could also be caused by the lack of precise knowledge of the parameter values
used in the model, there are two strong arguments against this model. First,
we should expect a variation in flux of at least a factor of 2 within the
phase interval of the BeppoSAX observations as follows: starting with a
gradual rise, peaking smoothly around phase 0.02 ($\sim5\times10^4$~s in
Fig.~\ref{lc}) and ending with a gradual decrease. Second, since the BeppoSAX
observation took place during periastron the X-ray luminosity should be the
largest, contrary to what it is observed (see Table~\ref{ew}).

Although it is clear that detailed observations covering a full orbital period
are needed in order to better study and model the orbital X-ray variability of
\object{LS~5039}, it seems unlikely that it can be responsible for the
variations of one order of magnitude quoted in Table~\ref{ew}. Therefore, we
conclude that this variability may be due to changes in the stellar wind
mass-loss rate on longer timescales.

%------------------------------------------------------------------------------
%------------------------------------------------------------------------------

\section{Conclusions} \label{conclusions}

The main conclusions of the analysis of our BeppoSAX observation of the
microquasar \lsrx\ can be summarized as follows:

\begin{enumerate}

\item We have obtained the most accurate value of the hydrogen column density
to date: $N_{\rm H}=1.0^{+0.4}_{-0.3} \times 10^{22}$~cm$^{-2}$. The derived
colour excess is compatible but higher than the one obtained from optical
photometry alone, suggesting that there might be intrinsic absorption
within the binary system orbit. 

\item Our observation took place near inferior conjunction, and no X-ray
eclipse has been detected. This fact has allowed us to compute upper limits
for the inclination of the binary system for different masses of the compact
object: $i<66\degr$ for $M_{\rm X}=1.4\msun$ and $i<68\degr$ for $M_{\rm
X}=9\msun$. These values are compatible with the ones obtained using the mass
function.

\item We have found hints of a correlation between the measured
$EW$(H$\alpha$) and the observed X-ray flux on timescales of years, in the
sense that the stronger the optical emission the higher the X-ray flux. In
this context, we have suggested that a decrease in the mass-loss rate from the
primary induces a decrease in the mass accretion rate onto the compact object,
and hence a decrease in the observed X-ray flux.

\item A simple Bondi-Hoyle accretion model (without taking into account a
varying absorption) cannot explain the X-ray variability because it predicts
an increase in the X-ray flux near periastron that is not observed. This may
happen because the model is very sensitive to some parameters not yet very
well constrained, like the eccentricity and the wind velocity. However, the
fact that no significant variability is detected near periastron suggests that
the observed variability in the X-ray flux on timescales of years is due to
changes in the mass-loss rate of the primary. An X-ray monitoring during a
full $\sim$4~d orbital cycle would clarify these ideas.

\end{enumerate}

%\appendix*{Orbital Bondi-Hoyle X-ray variability}
\section*{Appendix: Orbital Bondi-Hoyle X-ray variability}
%\label{app}

An order of magnitude estimate of the orbital X-ray variability can be
obtained using the Bondi-Hoyle accretion model. The accretion luminosity is
then computed by means of the simple formula:
\begin{equation}
L_{\rm X} \simeq \frac{G M_{\rm X} \dot{M}}{R_{\rm X}},
\label{eq:lx}
\end{equation}
where $G$ is the gravitational constant, $M_{\rm X}$ and $R_{\rm X}$ the mass
and radius of the compact object and $\dot{M}$ its variable accretion rate.
Assuming a neutron star, we will proceed adopting $M_{\rm X}=1.4\msun$
and $R_{\rm X}=10$ km. If the wind accretion onto the compact companion can be
described as a spherical inflow (Bondi \cite{bondi52}), the accretion rate as
a function of the binary separation $r$ is given by:
\begin{equation}
%\dot{M}=4\pi (G M_{\rm X})^2 \frac{\rho(r)}{v_{\rm rel}^3(r)},
\dot{M}=4\pi (G M_{\rm X})^2 \frac{v_{\rm w}(r) \rho(r)}{v_{\rm rel}^4(r)},
\label{eq:mdot}
\end{equation}
where $v_{\rm w}(r)$ is the wind velocity, $\rho(r)$ is the density of the
stellar wind and $v_{\rm rel}(r)$ its supersonical velocity relative to the
neutron star. We will approximate the wind velocity $v_{\rm w}(r)$ by a
typical $\beta$-law appropriate for O-type stars:
\begin{equation}
v_{\rm w}(r) = v_{\infty} \left[1- \frac{R_{\rm opt}}{r}\right]^{\beta}.
\label{eq:vw}
\end{equation}
Here, $v_{\infty}$ is the wind velocity at infinity and $\beta$ a power index.
Assuming a simple radial flow, the corresponding wind density can be derived
from the continuity equation as:
\begin{equation}
\rho(r) = \frac{\dot{M}_{\rm opt}}{4\pi r^2 v_{\rm w}(r)}.
\label{eq:rho}
\end{equation}
Following McSwain \& Gies (\cite{mcswain02}), suitable values for
\object{LS~5039} are $v_{\infty}=2850$ km~s$^{-1}$, $\beta=0.8$ and
$\dot{M}_{\rm opt}=8 \times 10^{-7}\msun$~yr$^{-1}$. To proceed, we have
solved numerically the Keplerian motion of the neutron star around the $M_{\rm
opt}=40\msun$ companion calculating both the eccentric and true anomalies
($E$ and $V$). The relative velocity between the wind and the neutron star is
then given by: 
\begin{eqnarray}
\lefteqn{v_{\rm rel}^2(r) = \left[\frac{n a^2}{r} \sin{E} + v_{\rm w}(r) \cos{V} \right]^2 + {} } \nonumber\\
& & {~~~~~~~~~}+\left[ \frac{n a^ 2 \sqrt{1-e^2}}{r} \cos{E} - v_{\rm w}(r) \sin{V} \right]^2,
\end{eqnarray}
where $a$ is the semimajor axis and $n=2\pi/P_{\rm orb}$ is the mean angular
motion.

\begin{acknowledgements}

We are grateful to J.~Miller for permission to include in this paper
unpublished results from his CHANDRA observation.
We thank useful comments and suggestions from an anonymous referee.
P.~R. is a researcher of the programme {\em Ram\'on y Cajal} funded by the
University of Valencia and Spanish Ministery of Science and Technology. M.~R.,
J.~M.~P. and J.~M. acknowledge partial support by DGI of the Ministerio de
Ciencia y Tecnolog\'{\i}a (Spain) under grant AYA2001-3092, as well as partial
support by the European Regional Development Fund (ERDF/FEDER). During this
work, M.~R. has been supported by a fellowship from CIRIT (Generalitat de
Catalunya, ref. 1999~FI~00199). J.~M. has been aided in this work by an Henri
Chr\'etien International Research Grant administered by the American
Astronomical Society, and has been partially supported by the Plan Andaluz de
Investigaci\'on of the Junta de Andaluc\'{\i}a (ref. FQM322). Skinakas
Observatory is a collaborative project of the University of Crete, the
Foundation for Research and Technology-Hellas and the Max-Planck-Institut
f\"ur Extraterrestrische Physik.

\end{acknowledgements}

\end{document}